\documentclass[showpacs,preprintnumbers,amsmath,amssymb,superscriptaddress]{revtex4-2} 
\usepackage{chemformula} 
\usepackage{graphicx}
\usepackage[colorlinks=true,citecolor=black,urlcolor=blue,linkcolor=black,bookmarksopen]{hyperref}
\usepackage{color}
\usepackage{braket}
\usepackage{physics}
\usepackage{gensymb}
\usepackage{soul}
\usepackage[T1]{fontenc}
\usepackage[utf8]{inputenc}
\usepackage[utf8]{inputenc}
\usepackage{setspace}
\usepackage{bibunits}

\usepackage{newunicodechar}
\newunicodechar{₂}{$_2$}

\usepackage[
  top=1in,
  bottom=1in,
  left=1.2in,
  right=1.2in
]{geometry}

\begin{document}

\title{Nonlinear Jackiw-Rebbi exciton-polaritons at van der Waals topological interfaces.}

\author{Oscar J. Palma Chaundler}
\email{ojpalmachaundler1@sheffield.ac.uk}
\author{Paul Bouteyre}
\thanks{Oscar J. P. Chaundler and P. Bouteyre contributed equally to this work as first authors.}
\author{Xuerong Hu}
\author{Sam A. Randerson}
\author{Daniel J. Gillard}
\author{Owen R. Wolley}
\affiliation{School of Mathematical and Physical Sciences, University of Sheffield, Sheffield S3 7RH, UK} 
\author{Yue Wang}
\affiliation{School of Physics, Engineering and Technology, University of York, York YO10 5DD, UK}

\author{Kenji Watanabe}
\affiliation{Research Center for Materials Nanoarchitectonics, National Institute For Materials Science, 1-1 Namiki, Tsukuba, 305-0044 Japan}
\author{Takashi Taniguchi}
\affiliation{Research Center for Materials Nanoarchitectonics, National Institute For Materials Science, 1-1 Namiki, Tsukuba, 305-0044 Japan}

\author{Helgi Sigurðsson}
\affiliation{Institute of Experimental Physics, Faculty of Physics, University
of Warsaw, ul. Pasteura 5, PL-02-093 Warsaw, Poland}

\author{Alexander I. Tartakovskii}
\email{a.tartakovskii@sheffield.ac.uk}
\affiliation{School of Mathematical and Physical Sciences, University of Sheffield, Sheffield S3 7RH, UK} 

\date{\today}

\begin{abstract}

Topological photonic structures provide a powerful means of confining and directing light through interface states formed between photonic phases with distinct topological properties. Van der Waals (vdW) materials offer complementary opportunities for nanophotonics: their optical properties can be selected from a wide family of layered crystals, while their atomically smooth surfaces and lack of lattice-matching constraints enable complex photonic heterostructures to be assembled layer by layer. This flexibility allows topological photonic confinement and strong excitonic coupling to be engineered separately within the same structure. Here, patterned low-loss dielectric hexagonal boron nitride (hBN) defines the photonic modes, while an encapsulated \ch{WSe_2} monolayer provides the excitonic resonance that couples directly to the confined optical field. We use this approach to realise Jackiw-Rebbi exciton-polaritons in a one-dimensional hBN double-grating heterostructure. Strong coupling between the \ch{WSe_2} exciton and a photonic Jackiw-Rebbi state localised at the interface between two topologically distinct gratings produces upper and lower Jackiw-Rebbi polariton branches. By varying the temperature from 4.7 to 300~K, we tune the exciton across the photonic bandgap and trace the evolution and anticrossing of these branches using angle-resolved reflectance and photoluminescence spectroscopy. Power-dependent photoluminescence further reveals a pronounced nonlinear response of the confined Jackiw-Rebbi polariton, including a blueshift of up to 5~meV and correlated changes in its intensity scaling and linewidth. These results establish vdW photonic heterostructures as a versatile platform for engineering tunable and nonlinear polaritons localised at topological interfaces.

\end{abstract}


\maketitle
\begin{bibunit}[apsrev4-2]
\section{Introduction}

Localised quantum mechanical states can appear at the boundary between two gapped systems with incompatible topological properties. Unlike conventional confined states arising from defects or localised potentials, these topologically confined states emerge because the global properties of the two systems do not match. This often leads to remarkably robust behaviour, forming the basis of the field of topological insulators \cite{Hasan2010}. Topological states have been studied extensively in electronic materials \cite{Hasan2010} and cold atoms in optical lattices \cite{Cooper2019}. More recently, analogous topological states have been discovered in photonics, showcasing a wide range of systems across the electromagnetic spectrum that host topologically protected optical modes \cite{Ozawa2019,Shalaev2019,parappurath2020}. 

In this work, we consider photonic gratings operating in the strong light-matter coupling regime~\cite{Zhang2018,Kravtsov2020}, which can support a class of topologically non-trivial states originally predicted as localised zero-energy solutions of the Dirac equation in the presence of a soliton field \cite{Jackiw1976}. These states, known as Jackiw-Rebbi (JR) interface states, are closely related, in the absence of spin, to the topologically protected Su-Schrieffer-Heeger (SSH) states discovered independently in linearly conjugated polymers \cite{Niemi1986}. Photonic analogues of both SSH chains \cite{Ozawa2019} and JR states \cite{Gorlach2019} have garnered considerable interest, with JR states exhibiting topological beaming \cite{Lee2022, choi2024} and hybridisation of interface modes \cite{Wang2025ArXiv}, offering potential applications in integrated optical routing and beam shaping. More recently, topological exciton-polaritons have also been demonstrated in perovskite junction metasurfaces \cite{An2024}.

In parallel with these developments, layered crystals known as van der Waals (vdW) materials have attracted tremendous interest over the last two decades due to their unique optical properties in the single- and few-layer limit. Strong exciton resonances with high quantum yields are exhibited throughout a diverse range of monolayers of 2D materials, particularly transition metal dichalcogenides (TMDs). As a result, monolayers of vdW materials have been integrated into a variety of silicon-, silicon nitride-, and III-V-semiconductor-based photonic and optoelectronic structures~\cite{Mak2016}. Their large exciton oscillator strengths and binding energies enable strong light-matter coupling with  a wide range of optical resonators, giving rise to exciton-polaritons in atomically thin semiconductors \cite{Schneider2018,Luo2024}. Since the first demonstrations of strong coupling in TMD monolayers embedded in optical microcavities \cite{Liu2015,Dufferwiel2015}, the field has rapidly expanded to include photonic-crystal structures \cite{Zhang2018}, optical bound states in the continuum \cite{Kravtsov2020}, and metasurface platforms \cite{Weber2023,Sortino2025}, providing a versatile toolbox for engineering polariton dispersion, transport, and nonlinear optical phenomena \cite{Luo2024}. Beyond neutral excitons, strong coupling has also been achieved with charged excitonic complexes (trions) \cite{Sidler2017,Emmanuele2020} and interlayer excitons~\cite{Datta2022,Louca2023}, revealing new opportunities to exploit many-body interactions and electronic tunability in polaritonic systems.

More recently, multilayer (quasi-bulk) vdW materials have also been explored as building blocks for nanophotonic devices. Like monolayers, these materials can be readily prepared by mechanical exfoliation and transferred onto a wide range of substrates without the need for chemical bonding or lattice matching \cite{Novoselov2005AtomicCrystals}. Their combination of high refractive index, low optical losses, and atomically smooth interfaces has established multilayer vdW materials as a versatile nanophotonic platform producing resonant metasurfaces, photonic crystals, Mie resonators, waveguides, and bound states in the continuum~\cite{Ermolaev2021,Munkhbat2022,Zotev2023,Weber2023,Zotev2025}. The availability of experimentally measured complex refractive index data for multilayer transition metal dichalcogenides and other vdW crystals has enabled accurate optical modelling and the design of all vdW photonic structures \cite{Ermolaev2021,Munkhbat2022,Zotev2023}. More recent work has expanded the materials palette to include anisotropic and polytype-engineered systems such as 3R-TMDs and rhombohedral boron nitride (rBN), which provide additional degrees of freedom for nonlinear optical processes, light confinement, and integrated photonic functionalities~\cite{Xu2022,Feng2024,Zograf2024,Trovatello2025}.

Building on these properties of vdW materials, spin-Hall topological edge-state propagation has been demonstrated in silicon-based structures incorporating TMD monolayers~\cite{Li2021}, and more recently in structures where quasi-bulk \ch{WS_2} serves directly as the photonic building block~\cite{Isoniemi2024}. Topological photonics has also been extended to the mid-infrared by integrating multilayer hBN with a silicon metasurface, where operation in the strong-coupling regime between topological photonic modes and hBN phonons enabled robust guiding of topological phonon-polaritons around sharp bends~\cite{Guddala2021}. These studies combined individual vdW flakes with conventional photonic structures or substrates, thereby only partially exploiting the versatility of vdW integration. More recently, we integrated nanostructured and planar layers of quasi-bulk \ch{WS_2} into a fully vdW photonic heterostructure supporting highly localised Jackiw-Rebbi interface states with directional far-field light reflection ~\cite{Randerson2026}. The integration of hBN-encapsulated monolayer \ch{WSe_2} into this multilayer architecture further enabled directional coupling of its excitonic emission to the photonic interface state.

Here, we combine topological photonics, exciton-polariton physics, and vdW heterostructure engineering to realise Jackiw-Rebbi exciton-polaritons in a fully vdW nanophotonic platform. Our 1D hBN double-grating heterostructure, with an embedded WSe$_2$ monolayer, supports a photonic Jackiw--Rebbi state that is strongly coupled to the \ch{WSe_2} exciton, forming light-matter states localised at the interface between two topologically distinct photonic phases. By tuning the exciton energy across the photonic bandgap through temperature, we directly resolve the formation and evolution of the upper and lower Jackiw-Rebbi polariton branches using angle-resolved reflectance and photoluminescence (PL) spectroscopy. We further show that these confined polaritons exhibit a pronounced nonlinear optical response, establishing vdW photonic heterostructures as a versatile platform for topological polaritonics and strongly interacting light-matter states.

\begin{figure*}[!ht]
\centering
\includegraphics[width=\linewidth]{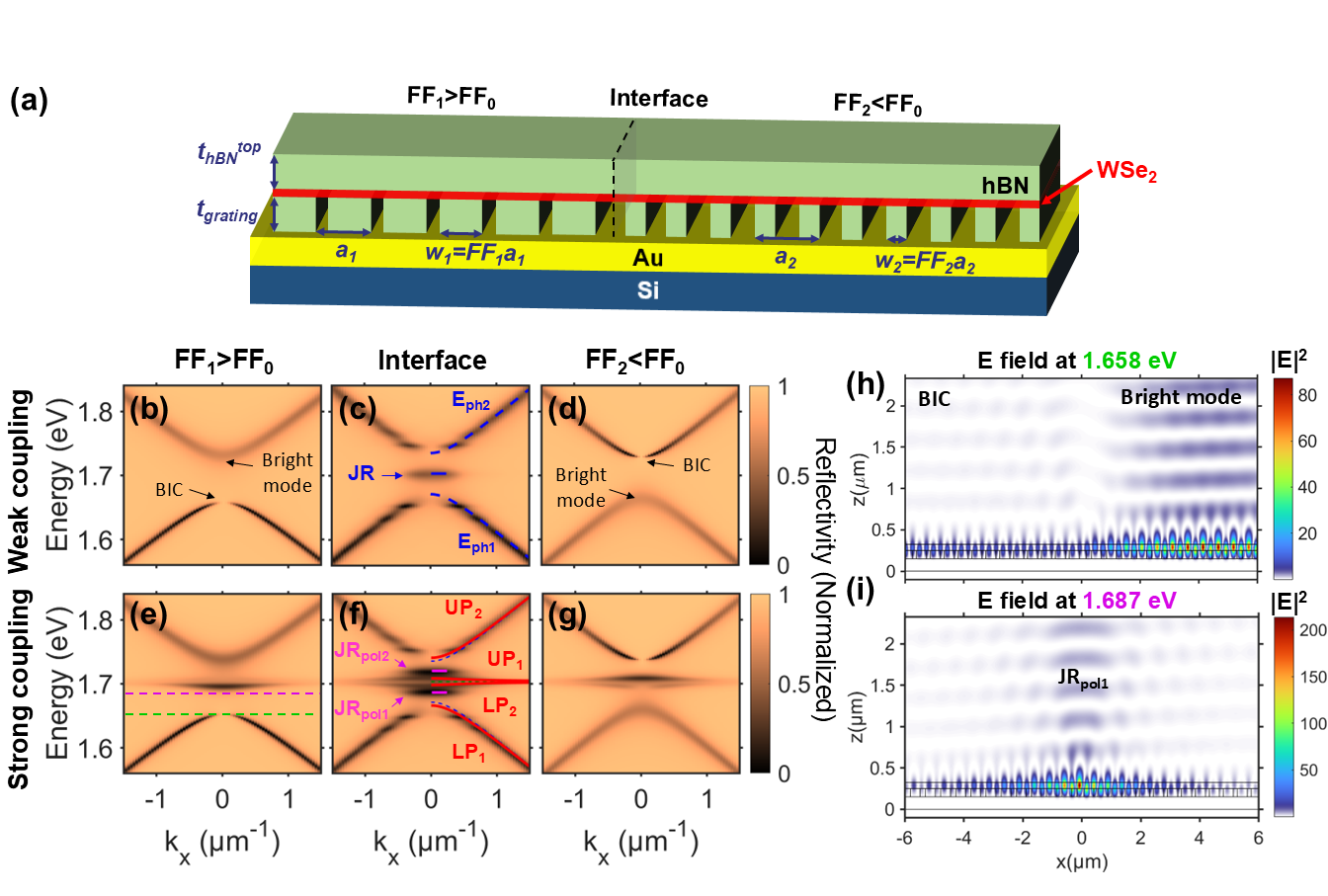}
    \caption{\textbf{Simulation of the \ch{hBN}/\ch{WSe_2} topological grating heterostructure.}
\textbf{a} Sketch of the \ch{hBN}/\ch{WSe_2} double-grating structure.
\textbf{b}--\textbf{d} Rigorous coupled-wave analysis (RCWA) simulations of the left grating ($a_1=480$ nm, $FF_1=0.7$) (b), right grating ($a_2=520$ nm, $FF_2=0.3$) (d), and the complete structure including the interface (c), without excitonic coupling. The blue dashed curves represent the photonic branches described by Eq.~(1), while the JR interface state is obtained from the full RCWA supercell calculation.
\textbf{e}--\textbf{g} Corresponding RCWA simulations of the left grating (e), right grating (g), and interface (f), incorporating the excitonic response of the \ch{WSe_2} monolayer at 220 K. The coloured curves represent the polariton energies obtained from the effective Hamiltonian in Eq.~(2), with red indicating the grating-mode polaritons and magenta indicating the JR interface-state polaritons. The analytical model uses a photonic midgap energy $E_0 = 1.70$ eV, coupling parameter $U = 31$ meV, group-velocity parameters $v_1 = v_2 = 0.085$ eV$\cdot\mu$m and bare JR-state energy $E_{\mathrm{JR}} = 1.70$ eV.
\textbf{h} Spatial distribution of the simulated electric-field intensity, $|E|^2$, at 1.658 eV, showing the bright and dark photonic modes.
\textbf{i} Spatial distribution of the electric-field intensity, $|E|^2$, at 1.687 eV, corresponding to the lower JR polariton.}
  \label{Figure1}
\end{figure*}

\section{Strong coupling of Jackiw-Rebbi edge state with \ch{WSe_2} excitons}
We realise JR exciton-polaritons using a photonic heterostructure approach developed in our recent work \cite{Randerson2026}. The \ch{WSe_2}/hBN topological double-grating heterostructure presented in Fig.~\ref{Figure1}(a) is composed of two 1D gratings etched next to one another in an \ch{hBN} flake of thickness $t_\text{grating}=80$ nm placed on a gold substrate. The stack of a \ch{WSe_2} monolayer and a planar  \ch{hBN} flake of thickness $t_\text{hbn}^\text{top}=80$ nm is then placed on top of the gratings. The gratings have periods $a_1=470$ nm and $a_2=550$ nm and filling factors (the ratio of the unetched beam to the period) $FF_1=0.73$ and $FF_2=0.29$, respectively. Here, $FF_0$ denotes the critical filling factor at which the photonic bandgap closes and the ordering of the symmetric and antisymmetric grating modes is inverted. 

We first perform rigorous coupled-wave analysis (RCWA) simulations with the \ch{WSe_2} monolayer retained in the structure but modelled using a constant background refractive index, without its excitonic resonance (see Methods for details). Here, the system is purely photonic, and figures \ref{Figure1} (b-d) show the corresponding angle-resolved reflectivity of individual gratings and the interface. The band structure of the TE-polarised modes within the first Brillouin zone consists of two branches with opposite curvatures, corresponding to symmetric and antisymmetric standing-wave photon modes. As functions of the wavenumber $k$ along the direction of the grating modulation, these modes are described by the following minimal Hamiltonian~\cite{Sigurdsson2024}:
\begin{equation}
	\label{eq1}
	\hat{H} =
	\left(\begin{matrix} 
	v k  &     Je^{i\phi}     \\
  Je^{-i\phi}    &   -v k      
 \end{matrix} \right)
 	-i \gamma \left(\begin{matrix} 
	1  &     1      \\
           1   &   1      
 \end{matrix} \right).
\end{equation}
Importantly, the Hamiltonian is non-Hermitian because it accounts for photon losses into the radiative continuum, predominantly through out-of-plane leakage. The second-order diffractive coupling, $Je^{i\phi}$, couples the counter-propagating guided TE modes within the \ch{hBN} slab, while $\gamma$ describes their first-order coupling to the radiative continuum and $v$ is their group velocity. A full derivation of this model is provided in Refs.~\cite{Randerson2026,Sigurdsson2024,Lu2020}. Here, $\hat{H}$ is expressed relative to the centre of the photonic bandgap.

The mirror symmetry of the grating far from the interface constrains $\phi$ to either $0$ or $\pi$, depending on the grating pitch and filling factor~\cite{Sigurdsson2024,Lu2020,Lee2021}. Consequently, the antisymmetric standing-wave photon mode supports a symmetry-protected photonic bound state in the continuum (BIC)~\cite{Azzam2021} with a theoretically infinite lifetime at the $\Gamma$ point. Such symmetry-protected BICs are commonly referred to as dark modes, as they are decoupled from the far-field radiative continuum at the $\Gamma$ point, in contrast to bright modes that couple efficiently to external radiation. For $\phi=0$, the BIC lies on the lower branch $\omega_-(k)$ (Fig.~\ref{Figure1}(b)), whereas for $\phi=\pi$, it lies on the upper branch $\omega_+(k)$ (Fig.~\ref{Figure1}(d)). Thus, changing the phase by $\pi$ exchanges the two photonic branches: the antisymmetric state hosting the BIC switches to the higher-energy branch, while the symmetric state switches to the lower-energy branch.

We now consider the interface between the two gratings, which share the same photonic bandgap and midgap energy but exhibit inverted ordering of the symmetric and antisymmetric photonic modes. As shown in Fig.~\ref{Figure1}(c), a topological interface state emerges within the bandgap as a nearly dispersionless horizontal feature in the simulated angle-resolved reflectivity map. This localised state arises from the topological incompatibility between the two gratings and is the photonic analogue of the Jackiw–Rebbi interface state originally predicted within the one-dimensional Dirac equation in relativistic quantum field theory ~\cite{Jackiw1976,Niemi1986}. A detailed description of the photonic origin of these mid-gap states is provided in our previous work~\cite{Randerson2026}.


The JR edge state exhibits topological properties, such as robustness against disorder, strong spatial localisation, and directional emission~\cite{Lee2022}, which may be retained when the state is strongly coupled to excitons in a \ch{WSe_2} monolayer, forming JR exciton-polaritons. This is particularly promising given the pronounced optical nonlinearities and effective particle interactions supported by exciton-polariton systems~\cite{Carusotto2013,Shahnazaryan_PRB2017,Kravtsov2020,Emmanuele2020,Datta2022,Louca2023}. We therefore incorporate an encapsulated WSe₂ monolayer into the double-grating heterostructure to couple its excitonic resonance directly to the JR interface state. We first investigate this coupling using RCWA simulations. The refractive index of the \ch{WSe_2} exciton is described by a Lorentz oscillator with an exciton energy $\hbar\omega_X = 1.7$ eV, a linewidth of 15 meV, and an amplitude of 1 eV$^2$, corresponding to parameters of the exciton resonance at a temperature of 220 K (see Methods for details). This temperature is chosen because the \ch{WSe_2} exciton energy lies near the centre of the photonic bandgap of the designed structure. 

Figures~\ref{Figure1}(e-g) show the simulated angle-resolved reflectivity maps of the two $WSe_2$ monolayer-coupled gratings, along with the topological interface between them. Strong coupling between the modes of each grating and the \ch{WSe_2} exciton gives rise to two sets of polaritons~\cite{Bouteyre2025}, referred to as upper (UP) and lower (LP) polaritons: (LP$_1$, UP$_1$) and (LP$_2$, UP$_2$). These branches are visible in the reflectivity maps in Figs.~\ref{Figure1}(e-g) and are labelled in red in Fig.~\ref{Figure1}(f). Importantly, the exciton also couples to the photonic JR edge state at the interface, giving rise to two JR exciton-polariton states, JR$_\text{pol1}$ and JR$_\text{pol2}$, labelled in pink in Fig.~\ref{Figure1}(f). Near the interface and for incident light at near-normal incidence ($\mathbf{k}_\parallel\approx 0$), this polaritonic system can be described by the following $6\times6$ Hamiltonian:

\begin{equation}
	\label{eq2}
	\hat{H}_{P}=
	\left(\begin{matrix} 
	\omega_- &  V  &   0    &   0    &    0    &  0     \\
      V  & \omega_X &   0    &   0    &    0    &  0     \\
        0   &   0    & \omega_\text{JR} &  V  &    0    &  0     \\
        0   &   0    &  V  & \omega_X &    0    &  0     \\
        0   &   0    &   0    &   0    & \omega_+ & V   \\
        0   &   0    &   0    &   0    &   V  & \omega_X \\
 \end{matrix} \right).
\end{equation}

We follow the exciton--photon spatial-overlap approximation introduced by Sigur{\dh}sson et al.~\cite{Sigurdsson2024}, treating the excitonic response as independent components coupled to the two photonic branches, $\omega_-$ and $\omega_+$, and to the JR interface state, with a common effective coupling strength $V$. The photonic branches $\omega_\pm$ are obtained from Eq.~(1) for the corresponding grating phase, $\phi = 0$ or $\pi$, while the bare JR-state energy is determined from the full interface calculation. Consequently, Eq.~(2) provides an effective description of the polariton energies but does not explicitly account for the spatial transition between the two gratings or the corresponding JR-state wavefunction, which are instead described by the RCWA supercell simulations. We define the zero of energy at the bare JR-state energy, such that $\omega_{\mathrm{JR}} = 0$.

The eigenvalues of this effective Hamiltonian are shown as coloured curves in Figs.~1(e--g), allowing direct comparison between the analytical polariton energies and the RCWA simulations. Further RCWA simulations illustrating the formation of these JR-polariton branches, together with a comparison with the analytical model, are provided in Supplementary Note~I.

The electric-field distributions of the complete topological structure, shown in Figs.~\ref{Figure1}(h) and~\ref{Figure1}(i), were calculated using RCWA at the energies of the bright and dark modes associated with LP$_2$, and of the lower JR polariton, JR$_\text{pol1}$, respectively. Here, $dark$ denotes a mode branch hosting a BIC that couples only weakly to the external radiative continuum. Figure~\ref{Figure1}(h) shows the field distribution at 1.658~eV, indicated by the green dashed line in Fig.~\ref{Figure1}(e). At this energy, the high- and low-filling-factor gratings support dark (BIC) and bright modes, respectively. The dark mode couples weakly to the radiative continuum in the high-filling-factor grating, whereas the bright mode couples efficiently to far-field radiation in the low-filling-factor region. 

Figure~\ref{Figure1}(i) shows the corresponding field distribution at 1.687~eV, indicated by the pink dashed line in Fig.~\ref{Figure1}(f), at the energy of the \textcolor{black}{lower} JR polariton, JR$_\text{pol1}$. The field is strongly localised at the interface between the two gratings while retaining efficient coupling to the far field. This spatially confined and directional far-field response can be understood as \emph{topological polariton beaming}, extending the concept of topological beaming previously demonstrated in purely photonic systems to a hybrid light-matter state~\cite{Lee2022,choi2024}. Such directional emission could enable improved light collection and controlled optical routing in future integrated polaritonic devices. \textcolor{black}{Further analysis of the spatial localisation of the JR-polariton mode, including its lateral decay along the x direction, vertical field distribution, and spectral dependence, is provided in Supplementary Note III.}

\begin{figure*}[ht!]
\centering
  \includegraphics[width=\linewidth]{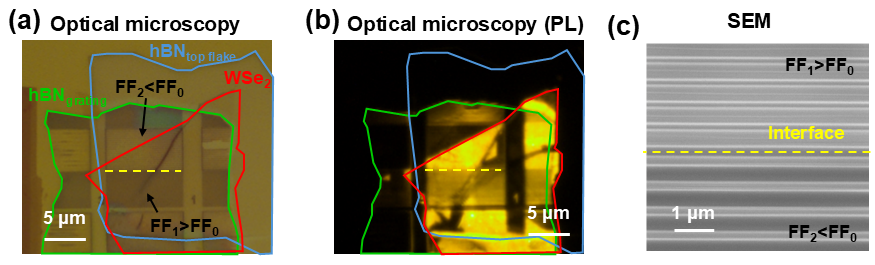}
\caption{\textbf{\ch{hBN}/\ch{WSe_2} topological grating heterostructure.}
\textbf{a} Bright-field microscopy image of the fabricated structure. The bottom patterned \ch{hBN} grating is indicated in green, the top \ch{hBN} flake in blue, and the \ch{WSe_2} monolayer in red. Multiple pairs of high- and low-filling-factor gratings are patterned within the same \ch{hBN} flake.
\textbf{b} Photoluminescence microscopy image of the heterostructure, showing emission from the encapsulated \ch{WSe_2} monolayer.
\textbf{c} Scanning electron microscopy (SEM) close-up image of the topological interface between the two grating regions. The dashed lines in (a) and (c) indicate the interface between the high- and low-filling-factor gratings.}
 \label{Figure2}
\end{figure*}

\begin{figure*}[ht!]
\centering
  \includegraphics[width=\linewidth]{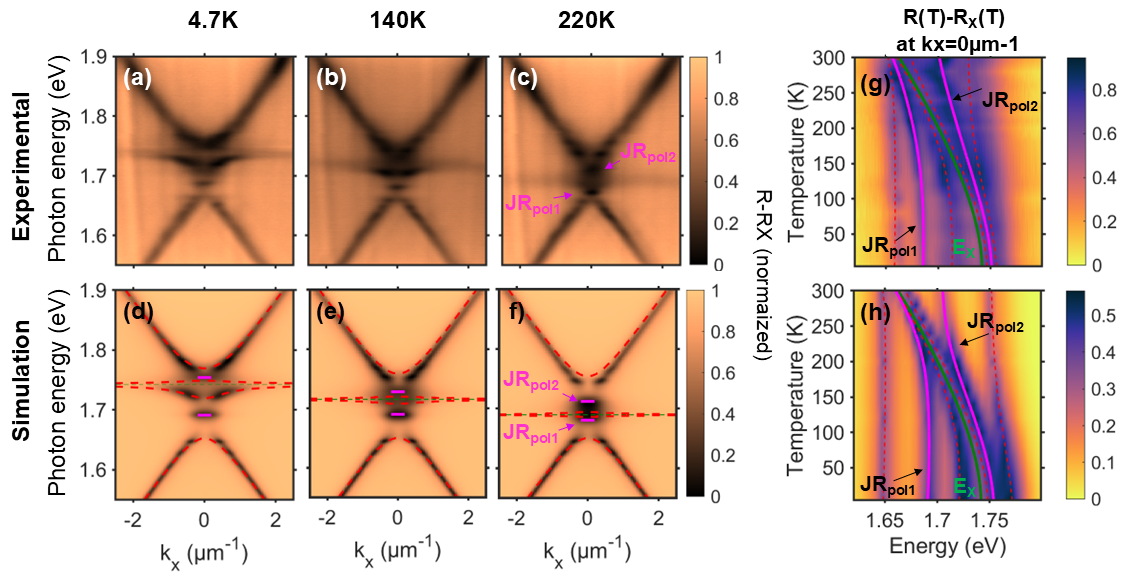}
\caption{\textbf{Temperature-dependent reflectivity of the \ch{hBN}/\ch{WSe_2} topological heterostructure.}
\textbf{a}--\textbf{c} Experimental angle-resolved reflectivity-contrast maps at 4.7, 140, and 220 K, respectively.
\textbf{d}--\textbf{f} Corresponding rigorous coupled-wave analysis (RCWA) simulations.
\textbf{g}--\textbf{h} Experimental and simulated temperature-dependent reflectivity-contrast maps, respectively, integrated around $k_x = 0~\mu\mathrm{m}^{-1}$.
The contribution from the uncoupled exciton, $R_X$, is subtracted from the reflectivity signal, $R$, to highlight the polaritonic modes.
The dashed curves indicate the grating-derived polariton branches, while the solid magenta curves identify the Jackiw--Rebbi (JR) interface-state polaritons, the green lines correspond to the exciton.}
 \label{Figure3}
\end{figure*}

\section{Temperature dependence}

So far, we have modelled the WSe₂ exciton-polariton system at 220 K, where the exciton energy lies near the centre of the photonic bandgap at 1.70 eV. We now experimentally investigate its temperature dependence from 4.7 to 300 K, focusing on the coupling of the exciton to the two grating modes and the JR interface state using the structure presented in Fig.~\ref{Figure2}. Figures~\ref{Figure2}(a) and (b) show bright-field and photoluminescence (PL) microscope images of the fabricated structure, respectively, while Fig.~\ref{Figure2}(c) shows a scanning electron microscopy (SEM) image of the topological interface. \textcolor{black}{The reproducibility of the inverted double-grating design and the realisation of the JR interface state are demonstrated with an additional heterostructure in Supplementary Note II.} 

Figure~\ref{Figure3} presents temperature-dependent angle-resolved reflectivity-contrast measurements obtained using a Fourier-imaging setup (see Methods for details). These measurements allow us to follow the evolution of exciton-photon coupling with temperature, paying particular attention to the JR interface state. Varying the temperature changes both the detuning and coupling strength between the \ch{WSe_2} monolayer exciton and the JR interface state. Specifically, the \ch{WSe_2} monolayer exciton energy redshifts with increasing temperature, following the Varshni relation~\cite{Varshni1967,Huang2016}, from 1.745~eV at 4.7~K to 1.67~eV at room temperature, while its oscillator strength decreases \textcolor{black}{(see Supplementary Information Notes IV and V for details).}
Figures~\ref{Figure3}(a-c) present the experimental reflectivity maps of the \ch{WSe_2}/hBN topological double-grating heterostructure at 4.7, 140, and 220~K, with the corresponding RCWA simulations shown in Figs.~\ref{Figure3}(d-f). \textcolor{black}{Reflectance contrast measurements at additional temperatures are presented in Supplementary Note IV.} The contribution from uncoupled exciton absorption, denoted $R_x$, has been removed from these maps to make the polaritonic dispersions more clearly visible (see Supplementary Note IV for details). The four remaining grating-mode polaritons (LP$_1$, UP$_1$, LP$_2$, and UP$_2$) and their dependence on detuning have been investigated in quasi-bulk WS$_2$ gratings in our recent work, where the photonic modes were coupled to the bulk exciton in WS$_2$~\cite{Bouteyre2025}. In the following, we focus primarily on the exciton-JR interface-state polaritons ($JR_\text{pol1}$ and JR$_\text{pol2}$).

At 4.7~K, the exciton energy is 1.745~eV in both the experiment and simulation (Figs.~\ref{Figure3}(a) and (d)). Due to the presence of band-to-band transitions above the exciton energy, the polaritonic features are less evident in this spectral region than below the exciton resonance. Accordingly, the lower JR interface-state polariton, JR$_\text{pol1}$, is visible in both experiment and simulation. The polaritonic model introduced previously in Eq.~\eqref{eq2} predicts that the upper JR interface-state polariton, JR$_\text{pol2}$, lies just above the exciton line. However, in this spectral region, the three polariton branches UP$_1$, UP$_2$, and JR$_\text{pol2}$ lie close together, making the upper JR interface-state polariton difficult to resolve (Fig.~\ref{Figure3}(d)).

At 140~K (Figs.~\ref{Figure3}(b) and~\ref{Figure3}(e)), the exciton redshifts to 1.72~eV, near the lowest energy point of the upper grating mode. At this temperature, the upper JR polariton, JR$_\text{pol2}$, becomes distinguishable in the simulated spectrum (Fig.~\ref{Figure3}(e)), although it is not resolved experimentally. The lower JR polariton, JR$_\text{pol1}$, also redshifts with the exciton in both the experiment and simulation. This spectral evolution is consistent with exciton–photon hybridisation and suggests the presence of strong coupling between the photonic JR state and the \ch{WSe_2} exciton. 

At 220~K (Figs.~\ref{Figure3}(c) and~\ref{Figure3}(f)), the exciton redshifts to 1.70~eV, bringing it into resonance with the midgap energy of the structure considered in Fig.~\ref{Figure1}. As expected, both the lower and upper JR polaritons, JR$_\text{pol1}$ and JR$_\text{pol2}$, are resolved in the experimental and simulated spectra. The energy separation between these branches near resonance, together with their temperature-dependent evolution, further supports their hybrid light–matter character.

Finally, the topological polaritonic system at 300~K, shown in Supplementary Note IV, represents nearly the opposite detuning regime to that at 4.7~K. At this temperature, the exciton has redshifted to 1.67~eV, approaching the energy of the lower grating mode and reaching negative detuning relative to the JR state. However, the exciton oscillator strength is lower, and its linewidth is broader, resulting in weaker light-matter coupling. In this case, only the upper JR interface-state polariton, JR$_\text{pol2}$, is visible in both experiment and simulation. As observed for the lower JR interface-state polariton at lower temperatures, the upper JR interface-state polariton shifts with the exciton energy, providing further evidence of light-matter hybridisation. 

Figures~\ref{Figure3}(g) and (h) show the experimental and simulated reflectivity maps, respectively, integrated around $k_x=0~\mu\mathrm{m}^{-1}$ and plotted as functions of temperature for all measurements, including those presented in \textcolor{black}{Supplementary Note IV}. The contribution from uncoupled exciton absorption has again been removed from these temperature-dependent reflectivity maps (see Methods for details). These maps reveal the temperature dependence of the individual polariton modes in finer detail. Each mode energy is fitted using the polaritonic model presented in Eq. ~\eqref{eq2}, accounting for the temperature dependence of both the exciton energy and the coupling strength (see Supplementary Note IV). The JR interface-state polaritons, shown in pink, exhibit a clear anticrossing with the \ch{WSe_2} exciton, demonstrating strong coupling between the exciton and the JR interface state. \textcolor{black}{To further quantify the strong-coupling regime, we analyse the reflectance spectra and extract the energies and linewidths of the JR polariton and excitonic resonances (Supplementary Note V). The fitted JR-polariton branches present an avoided crossing as the exciton is tuned in energy. Meanwhile, the Hopfield coefficients show the continuous exchange of photonic and excitonic character. Using the extracted excitonic linewidth, the linewidth of the predominantly photonic JR-polariton branch as an estimate of the photonic linewidth, and the Rabi splitting obtained from the coupled-oscillator model, we evaluate the strong-coupling ratio \(R_{\mathrm{SC}}=\hbar\Omega_R/[(\gamma_{\mathrm{ex}}+\gamma_{\mathrm{ph}})/2]\). As seen in Supplementary Note V, this ratio remains above unity from cryogenic temperatures up to 260 ~K, showing that the JR interface state remains in the strong-coupling regime over most of the temperature range investigated.}

\begin{figure*}[!ht]
\centering
\includegraphics[width=\linewidth]{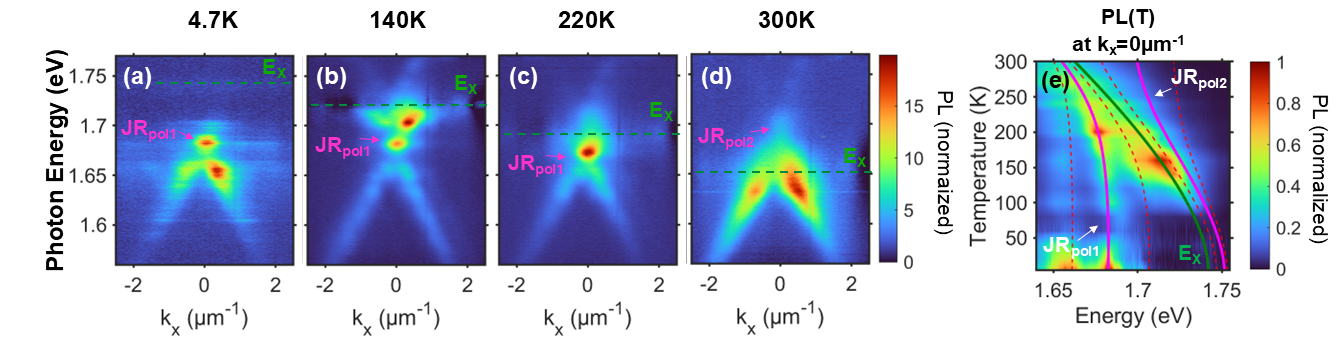}
\caption{\textbf{Temperature-dependent photoluminescence of the \ch{hBN}/\ch{WSe_2} topological heterostructure.}
\textbf{a}--\textbf{d} Experimental angle-resolved photoluminescence (PL) maps of the heterostructure at 4.7, 140, 220, and 300 K, respectively. Emission from the lower Jackiw--Rebbi (JR) polariton, $\mathrm{JR}_{\mathrm{pol1}}$, is observed up to 220 K, while a weak contribution from the upper JR polariton, $\mathrm{JR}_{\mathrm{pol2}}$, becomes visible at 300 K.
\textbf{e} Temperature-dependent PL intensity integrated around $k_x = 0~\mu\mathrm{m}^{-1}$ for all measured temperatures, showing the evolution of the uncoupled exciton and JR-polariton emission. The overlaid curves indicate the temperature evolution of the exciton and polariton energies.}
  \label{Figure4}
\end{figure*}

\section{Photoluminescence of the Jackiw-Rebbi polaritons} 

We next investigate the temperature-dependent PL of the JR polaritons under excitation with a pulsed supercontinuum laser (see Methods for details). Figure~\ref{Figure4} shows the angle-resolved PL from the \ch{hBN}/\ch{WSe_2} topological heterostructure at 4.7, 140, 220, and 300~K (Figs.~\ref{Figure4}(a-d), respectively), corresponding to the temperatures examined above by reflectivity spectroscopy. 

At all temperatures, emission from the uncoupled \ch{WSe_2} monolayer exciton and the lower grating-mode polariton, LP$_1$, is clearly visible. At 4.7, 140, and 220~K (Figs.~\ref{Figure4}(a-c)), the spectra also show pronounced emission from the lower JR polariton, JR$_\text{pol1}$. At 300~K (Fig.~\ref{Figure4}(d)), emission from this branch is no longer resolved, while a weak signal associated with the upper JR polariton, JR$_\text{pol2}$, becomes visible. This evolution is consistent with the temperature-dependent changes in exciton-photon detuning observed in the reflectivity measurements of Fig. \ref{Figure3}. 

Figure~\ref{Figure4}(e) shows the photoluminescence intensity integrated around $k_x=0~\mu\mathrm{m}^{-1}$ as a function of temperature. The map reveals the temperature evolution of the uncoupled \ch{WSe_2} exciton and the lower JR polariton, JR$_\text{pol1}$, whose emission follows the exciton redshift over a broad temperature range. 

These measurements demonstrate efficient population of, and radiative emission from, the lower JR polariton, confirming that the strongly coupled state identified in reflectivity is also observable in PL.

\section{Nonlinearities of Jackiw-Rebbi polaritons}
\begin{figure*}[ht!]
\centering
\includegraphics[width=\linewidth]{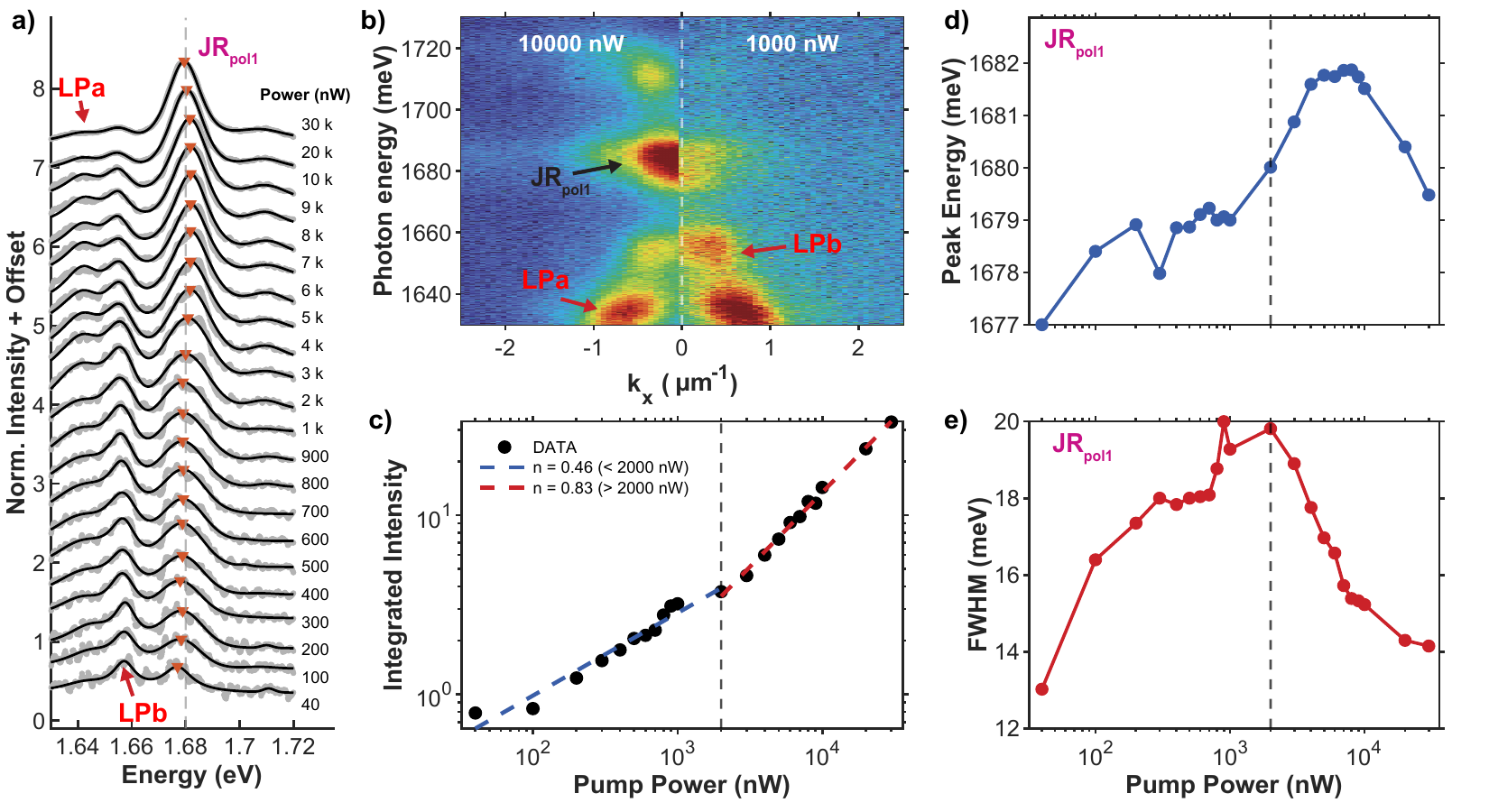}
\caption{\textbf{Power-dependent photoluminescence of the \ch{hBN}/\ch{WSe_2} topological heterostructure at 4.7 K.}
\textbf{a} Experimental PL spectra integrated around $k_x = 0~\mu\mathrm{m}^{-1}$ for excitation powers ranging from 40 to 30,000 nW. The spectra are vertically offset for clarity, highlighting the evolution of the grating-derived polariton modes and the lower JR polariton, $\mathrm{JR}_{\mathrm{pol1}}$.
\textbf{b} Angle-resolved PL maps measured at 10,000 nW (negative $k_x$) and 1,000 nW (positive $k_x$), with each half independently normalised to highlight changes in the relative emission intensity of the polariton branches.
\textbf{c} Integrated intensity of $\mathrm{JR}_{\mathrm{pol1}}$ as a function of excitation power. Dashed lines represent power-law fits, $I \propto P^n$, yielding $n = 0.46$ below and $n = 0.83$ above approximately 2,000 nW.
\textbf{d} Fitted peak energy of $\mathrm{JR}_{\mathrm{pol1}}$ as a function of excitation power, showing an initial blueshift followed by a redshift at the highest powers.
\textbf{e} Corresponding full width at half maximum (FWHM) of the JR-polariton emission, showing an initial broadening followed by linewidth narrowing. The vertical dashed lines in (c)--(e) indicate the characteristic excitation-power crossover near 2,000 nW.}
    \label{Figure5}
\end{figure*}

To investigate the nonlinear response of the JR interface state, we performed a systematic power-dependent PL study at the grating interface illustrated in Fig.~\ref{Figure1}(a), with the resulting spectra and extracted parameters presented in Fig. ~\ref{Figure5} . The heterostructure was excited using a pulsed ($20$~ps) broadband supercontinuum laser covering the wavelength range from 587 to 687~nm, with a repetition rate of 152~kHz. The excitation power was varied over almost three orders of magnitude, from 40 to 30,000~nW. \textcolor{black}{ Our analysis focuses on the PL cross-section at $k_x=0$, using a multi-Voigt fitting procedure to distinguish the JR-polariton emission from the overlapping lower polariton bands. Representative spectra and their corresponding fits are presented in Supplementary Note VI. Each spectrum was decomposed into three contributions: $\mathrm{LPB}_a$ and $\mathrm{LPB}_b$, associated with the slightly different $\mathrm{LPB}_1$ resonances of the two grating regions and the lower JR interface polariton  JR$_\text{pol1}$. Here, the labels a and b distinguish the two grating-derived polariton contributions resolved in the PL spectra. This decomposition allows us to extract the peak energy, linewidth, and integrated intensity of the interface state.} The power dependence of these extracted parameters is summarised in Fig.~\ref{Figure5}. \textcolor{black}{The waterfall plot in Fig.~\ref{Figure5}(a) presents an overview of the spectral evolution across the full range of excitation powers, while representative multi-Voigt fits and their individual spectral contributions are presented in Supplementary Fig. S6(a). The left and right panels of Fig.~\ref{Figure5}(b) show normalised PL dispersion maps measured at high (10,000~nW) and moderate (1000~nW) excitation powers, respectively. Normalised PL is shown for each individual panel. The normalised spectra reveal a substantial change in the relative intensity of the JR interface-state flat band, accompanied by a blueshift: the maximum PL signal shifts from the LP$_1$ band to JR$_\text{pol1}$.}

Analysis of the power-dependent response reveals a clear change in the scaling of the integrated intensity at approximately 2000~nW in Fig.~\ref{Figure5}(c), where the power-law exponent increases from $n=0.46$ to $n=0.83$. Below 2000~nW, the strongly sub-linear dependence may reflect inefficient conversion of photoexcited carriers into JR polaritons, with carriers also scattering into the LP$_1$ and LP$_2$ branches. Above 2000~nW, the increased relative intensity of the JR polariton peak suggests that the interface state receives a larger fraction of the photoexcited population. Similar excitation-dependent redistribution and nonlinear scattering between polariton states have been reported in semiconductor microcavities~\cite{Ballarini2013,Zhao2022}. In the present system, however, the integrated JR-polariton intensity exhibits sublinear scaling both below and above 2000 nW, and the observed crossover does not by itself establish stimulated parametric scattering.

The JR polariton peak energy as a function of excitation power is shown in Fig.~\ref{Figure5}(d). At low and intermediate powers, the peak blueshifts continuously from approximately 1677 to 1682~meV. This excitation-power-dependent blueshift is consistent with repulsive interactions between photoexcited quasiparticles (including polariton–polariton interactions and interactions with the photoexcited exciton reservoir). Comparable blueshifts have been reported in III-V microcavity polaritons and in TMD polariton systems, including monolayer \ch{MoSe_2} coupled to optical bound states in the continuum~\cite{Kravtsov2020} and monolayer \ch{WS_2} coupled to Bloch surface waves~\cite{Barachati2018}, as well as in pump-induced polariton potentials in III-V semiconductor microcavities~\cite{Sanvitto2006,Wertz2010}.

Towards the highest excitation powers in Fig. ~\ref{Figure5} (d), the blueshift partially reverses and the JR-polariton emission peak moves towards lower energies. Although this behaviour suggests a change in the mechanisms governing the polariton energy shift, its microscopic origin cannot be established from the present measurements. In particular, the integrated PL intensity continues to increase without clear evidence of saturation (Fig. ~\ref{Figure5}(c)), preventing an unambiguous attribution of the observed redshift to exciton saturation. 

The full width at half maximum (FWHM) of the JR polariton emission is shown as a function of excitation power in Fig.~\ref{Figure5}(e). At low powers, the linewidth broadens from approximately 13~meV to a maximum of 19-20~meV near the characteristic excitation power of 2000~nW. This initial broadening is consistent with increased scattering involving polaritons, excitons, and the photoexcited reservoir. Above 2000~nW, the linewidth narrows rapidly to approximately 14~meV. Linewidth narrowing at elevated excitation densities has also been observed in monolayer-TMD polariton systems and is commonly associated with changes in polariton relaxation, occupation, and coherence~\cite{Waldherr2018}. Here, the simultaneous linewidth narrowing and increase in the intensity exponent suggest that both features arise from a common change in the population and relaxation dynamics of the confined JR polariton. However, because the intensity remains sub-linear, we do not interpret this crossover as evidence of polariton condensation or lasing. 

Taken together, the power-dependent evolution of the integrated intensity, peak energy, and linewidth demonstrates that the confined JR polariton supports a pronounced nonlinear optical response. The blueshift of approximately 5~meV is consistent with repulsive interactions, while the concurrent changes in intensity scaling and linewidth reveal a crossover in the population and relaxation dynamics near 2000~nW. \textcolor{black}{The spatial field distribution will play an important role in the nonlinear response. The encapsulation of the \ch{WSe_2} monolayer within the hBN heterostructure places the exciton directly within the confined JR-polariton field, providing an overlap between the exciton and the optical mode (Supplementary Note III). However, the calculated field maximum lies predominantly on the hBN slab rather than at the monolayer. Further optimisation of the heterostructure geometry, aimed at shifting the mode-field maximum towards the $WSe_2$ monolayer, could increase the light–matter overlap and provide a route towards stronger nonlinear interactions.} The combination of topological interface localisation, the excitonic nonlinearity of monolayer \ch{WSe_2}, and the high-quality dielectric environment provided by hBN establishes vdW photonic-crystal gratings as a promising platform for nonlinear topological polaritonics.

\section{Conclusion} 

We have realised Jackiw-Rebbi exciton-polaritons in a fully vdW hBN/\ch{WSe_2} photonic heterostructure. In this architecture, patterned hBN defines a photonic JR state localised at the interface between two topologically distinct gratings, while an encapsulated \ch{WSe_2} monolayer provides the excitonic resonance. Strong coupling between these components produces upper and lower JR polariton branches confined at the topological interface. 

Temperature tuning of the \ch{WSe_2} exciton from 4.7 to 300~K allowed us to control its detuning from the JR state and directly resolve the resulting anticrossing using angle-resolved reflectance spectroscopy. The temperature evolution of the polariton branches is reproduced by RCWA simulations and an analytical coupled-oscillator model. Photoluminescence measurements further demonstrate efficient population and radiative emission of the JR polaritons over a broad temperature range. 

Power-dependent photoluminescence reveals that the confined JR polariton supports a pronounced nonlinear optical response, including a blueshift of approximately 5~meV and correlated changes in its intensity scaling and linewidth near an excitation power of 2000~nW. Together, these results establish vdW photonic heterostructures as a powerful platform in which topological confinement, light-matter coupling, and optical nonlinearities can be engineered within the same structure. This approach opens a route towards compact and tunable topological polariton systems for nonlinear nanophotonics and the study of interacting light-matter states.

\section*{Acknowledgements}
O.J.P.C., P.B., X.H., S.A.R., D.J.G., O.R.W. and A.I.T. thank EPSRC grants EP/V006975/1, EP/V007696/1, EP/V026496/1, EP/S030751/1. Yue Wang thanks EPSRC grant EP/V047663/1 and Royal Academy of Engineering fellowship RF/201718/17131. H.S. acknowledges the National Science Center, Poland, project No. 2024/55/B/ST3/02954.

\section*{Methods}
\subsection{Simulations}

One-dimensional (1D) van der Waals (vdW) grating structures were simulated using $S^4$, an open-source implementation of the rigorous coupled-wave analysis (RCWA) method with a Python interface. To model each grating region, a unit cell with a specified periodicity was defined for the multilayer vdW heterostructure. The model included from bottom to top a Si/Au substrate, an hBN grating, a \ch{WSe_2} monolayer, and a top hBN slab. The thickness of each layer was specified, and the unit cell was repeated periodically to represent an infinite 1D grating. 

\textcolor{black}{The optical properties of hBN were taken from Ref. \cite{Zotev2023}. The optical response of the \ch{WSe_2} monolayer was described using a Lorentz oscillator. For the simulations at 220 K, the exciton energy was set to 1.70 eV, with a linewidth of 15 meV and an oscillator amplitude of 1 eV². For the temperature-dependent simulations, the exciton resonance energy was varied according to the Varshni relation, while the oscillator strength was adjusted to account for its temperature dependence. The resulting modification of the complex refractive index and the formation of the exciton-polariton branches are illustrated in Supplementary Note I.} The grating layer was patterned with a specified filling factor and period to achieve the desired periodic modulation. Simulations were performed for TE polarisation of incident light by varying the angle of incidence in the plane perpendicular to the grating grooves.

Photonic band inversion was achieved by tuning the filling factor and period of the individual 1D gratings. To reproduce the JR interface state between the two grating regions, a supercell was defined that was periodically repeated in simulations in the infinite structure. This supercell contained 40 grooves in total: 20 grooves with the period and filling factor corresponding to the topological regime, in which the BIC lies on the lower photonic band, followed by 20 grooves with the period and filling factor corresponding to the trivial regime, in which the BIC lies on the upper photonic band. The resulting supercell supports a JR state localised at the interface between the two grating regions. The lateral size of the supercell ($\sim 20~\mu$m) exceeds the spatial extent of the JR state ($\sim 10~\mu$m, including its evanescent tails; see Fig.~S3), minimising artificial coupling between JR states in neighbouring periodically repeated supercells. The RCWA calculations were performed using $N_{\mathrm{ord}} = 100$ Fourier harmonics.

\subsection{Inverted Grating Heterostructure Fabrication}

\ch{hBN} crystals provided by K. Watanabe and T. Taniguchi were mechanically exfoliated onto Si substrates coated by electron-beam evaporation with 150/10~nm gold/titanium film. The substrates were maintained at 65~\degree C during exfoliation to promote flake adhesion. After identifying \ch{hBN} flakes with the desired thickness of approximately 100~nm (using atomic-force microscopy, AFM), a positive electron-beam resist (ARP-13, AllResist GmbH) was spin-coated at 4000~rpm for 30~s and baked at 180~\degree C for 2~min. 

Electron-beam lithography was performed using a Raith Voyager system operated at an accelerating voltage of 50~kV and a beam current of 1.2~nA. Multiple pairs of gratings with high and low filling factors were patterned on the same \ch{hBN} flakes to produce the required photonic band inversion. To account for variations in the etching process, the grating pairs were exposed using a range of electron doses. The patterns were then transferred into the \ch{hBN} using reactive-ion etching for 40~s at a pressure of 0.14~mbar and a DC bias of 135~V, with a combined chemical and physical etching process based on \ch{SF_6} and \ch{CHF_3} gases. This process produced parallel \ch{hBN} grating bars without preferential etching along the crystallographic zigzag direction. 

As shown in Fig.~\ref{Figure2}(a), the fabricated structure contains two adjoining grating regions. The light region, labelled FF2, has the lower filling factor, while the darker region, labelled FF1, has the higher filling factor. The filling factor is defined as the width of the unetched \ch{hBN} bar divided by the grating period. The two filling factors were selected to place the corresponding gratings on opposite sides of the photonic band inversion and achieve similar photonic bandgaps\cite{Randerson2026}. At the critical filling factor, FF0, the photonic bandgap closes at a pair of exceptional points~\cite{Randerson2026}. The grating thickness, filling factors and periods were selected with an account of the top \ch{hBN} slab added at the final fabrication step.

Figure~\ref{Figure2}(c) shows a scanning electron microscopy (SEM) image of the grating interface, from which the fabricated periods and filling factors were extracted. Grating 1 had a period of 470~nm and a filling factor of 0.73, while grating 2 had a period of 550~nm and a filling factor of 0.29. The parameters extracted from the SEM images were used in RCWA simulations to determine the top \ch{hBN} slab thickness required to align the photonic bandgaps of the two gratings while maintaining their inverted band ordering. The target thickness of the top \ch{hBN} slab was 74~nm. 

A polymer membrane was then prepared for assembly of the vdW heterostructure. First, a sacrificial layer of PMGI was spin-coated onto a Si substrate at 8000~rpm for 2~min and baked at 130~\degree C for 5~min. A layer of PMMA was subsequently spin-coated over the PMGI at 3500~rpm for 2~min, with the spin speed selected to obtain the required membrane thickness, followed by a second bake at 130~\degree C for 5~min. The top \ch{hBN} slab was prepared by mechanically exfoliating \ch{hBN} onto the PMMA membrane at 65~\degree C. AFM was used to identify a flake with a thickness of 74~nm and sufficient lateral dimensions to cover the patterned grating regions. A circle approximately 3~mm in diameter was then scribed around the selected flake. The underlying PMGI layer was dissolved in MF-319 developer, after which the substrate was immersed in deionised water to release the floating PMMA membrane. The membrane was collected using a cantilever coated with PMMA to provide an adhesive surface. The cantilever and membrane were then baked at 150~\degree C for 5~min to improve adhesion. The membrane was prepared under cleanroom conditions. 

A \ch{WSe_2} monolayer was encapsulated between the patterned bottom \ch{hBN} layer and the planar top \ch{hBN} slab, as shown in Fig.~\ref{Figure2}(a). The \ch{WSe_2} was mechanically exfoliated onto polydimethylsiloxane (PDMS), and a monolayer of sufficient lateral dimensions was identified by optical microscopy. PDMS was used as the exfoliation substrate because it facilitated the isolation and subsequent transfer of large monolayer flakes with lateral sizes exceeding 20 $\mu$m.  The heterostructure was assembled inside a nitrogen-filled glovebox. Using a robotic micromanipulator and an optical microscope, the top \ch{hBN} slab supported by the PMMA membrane was aligned with the \ch{WSe_2} monolayer. The monolayer was picked up at 30~\degree C, forming an \ch{hBN}/\ch{WSe_2} stack. This stack was then aligned with the patterned \ch{hBN} gratings on the gold-coated substrate and released at 65~\degree C. Finally, the PMMA membrane was removed using acetone, followed by rinsing with isopropyl alcohol. 

Figure~\ref{Figure2}(a) shows the resulting heterostructure, with the \ch{WSe_2} monolayer and top \ch{hBN} slab positioned over the target grating interface. No large trapped bubbles are visible within the optically active region. PL imaging (Fig.~\ref{Figure2}(b)) confirms that the encapsulated \ch{WSe_2} monolayer retains its bright emission after assembly and solvent cleaning. This fabrication procedure therefore enables the deterministic assembly of the inverted \ch{hBN} grating heterostructure while preserving the optical quality of the embedded monolayer.

\subsection{Temperature-dependent angle-resolved reflectance} 

Temperature-dependent angle-resolved reflectance-contrast measurements were performed using a home-built Fourier-imaging spectroscopy setup described previously \cite{Bouteyre2025,Randerson2026}. The sample was mounted under vacuum on a fixed stage in a liquid-helium flow cryostat. The stage was thermally coupled to a cold finger through which liquid helium was circulated and was equipped with a resistive heater. The sample temperature was controlled between 4.7~K and room temperature by adjusting the helium flow and heater power. The sample was illuminated with a broadband white-light source (Thorlabs, SLS201L), and the reflected light was collected through a microscope objective of numerical aperture 0.7. The back focal plane of the objective was imaged onto the entrance slit of a spectrometer, allowing the photon energy and in-plane wavevector perpendicular to the grating grooves to be measured simultaneously. An intermediate imaging stage equipped with an adjustable rectangular aperture allowed selection of a specific area on the grating for which the angular dependence was eventually measured by the spectrometer/CCD system. Measurements were performed under TE polarisation, with the electric field oriented parallel to the grating grooves. Reflectance contrast was calculated as
\begin{equation}
    RC = \frac{R_{\mathrm{sample}} - R_{\mathrm{ref}}}
    {R_{\mathrm{ref}}},
\end{equation}
where $R_{\mathrm{sample}}$ is the reflectance measured
from the selected grating region and $R_{\mathrm{ref}}$
is the reflectance measured from a reference region
of the sample.

\subsection{Angle-resolved photoluminescence} 

For all PL measurements, the sample was mounted in the same liquid-helium flow cryostat and measured using the same Fourier-imaging detection setup as for the reflectance-contrast measurements. The broadband white-light source was replaced with a picosecond supercontinuum laser with a central wavelength and bandwidth as given below, and a 700~nm long-pass filter was placed in the collection path to suppress scattered excitation light. Temperature-dependent PL measurements were performed using an average excitation power of 200~$\mu$W, a spectral bandwidth of 10~nm centred at 637~nm, and a repetition rate of 78~MHz. Angle-resolved emission was collected over the same momentum direction as in the reflectance measurements. For the power-dependent measurements, the sample was excited using a 100~nm spectral bandwidth extending from 587 to 687~nm and centred at 637~nm, with a repetition rate of 152~kHz. The average excitation power was varied from 40 to 30,000~nW. The resulting spectra were analysed at $k_x=0~\mu\mathrm{m}^{-1}$, as described in Supplementary Note~VI.

\section*{Data availability}

The data that support the findings of this study are available from the corresponding author upon reasonable request.

\putbib[bib]
\end{bibunit}
\clearpage
\renewcommand{\thefigure}{S\arabic{figure}}
\renewcommand{\thetable}{S\arabic{table}}
\renewcommand{\theequation}{S\arabic{equation}}
\renewcommand{\theHfigure}{S\arabic{figure}} 
\renewcommand{\theHtable}{S\arabic{table}}
\renewcommand{\theHequation}{S\arabic{equation}}
\setcounter{figure}{0}
\setcounter{table}{0}
\setcounter{equation}{0}
\setcounter{section}{0}
\renewcommand{\thesection}{\Roman{section}}

\begin{center}
\textbf{\large Supplementary Information for\\
Nonlinear Jackiw-Rebbi exciton-polaritons at van der Waals topological interfaces}
\end{center}
\begin{center}

\vspace{0.5cm}

{\normalsize
Oscar J. Palma Chaundler$^{1,*}$,
Paul Bouteyre$^{1,\dagger}$,
Xuerong Hu$^{1}$,
Sam A. Randerson$^{1}$,
Daniel J. Gillard$^{1}$,
Owen R. Wolley$^{1}$,
Yue Wang$^{2}$,
Kenji Watanabe$^{3}$,
Takashi Taniguchi$^{3}$,
Helgi Sigur{\dh}sson$^{4}$,
and Alexander I. Tartakovskii$^{1,\ddagger}$
\par}

\vspace{0.3cm}

{\small\itshape
$^{1}$School of Mathematical and Physical Sciences,\\
University of Sheffield, Sheffield S3 7RH, UK\\
$^{2}$School of Physics, Engineering and Technology,\\
University of York, York YO10 5DD, UK\\
$^{3}$Research Center for Materials Nanoarchitectonics,\\
National Institute for Materials Science,\\
1-1 Namiki, Tsukuba, 305-0044 Japan\\
$^{4}$Institute of Experimental Physics, Faculty of Physics,\\
University of Warsaw, ul. Pasteura 5,\\
PL-02-093 Warsaw, Poland
\par}

\end{center}

\vfill

{\footnotesize
\noindent\rule{4cm}{0.4pt}\\
$^{*}$ ojpalmachaundler1@sheffield.ac.uk\\
$^{\dagger}$ Oscar J. P. Chaundler and P. Bouteyre
contributed equally to this work as first authors.\\
$^{\ddagger}$ a.tartakovskii@sheffield.ac.uk
}
\begin{bibunit}[apsrev4-2]
\section{Supplementary Note I}
\label{sec:S1}

\begin{figure*}[ht!]
\centering
  \includegraphics[width=\linewidth]{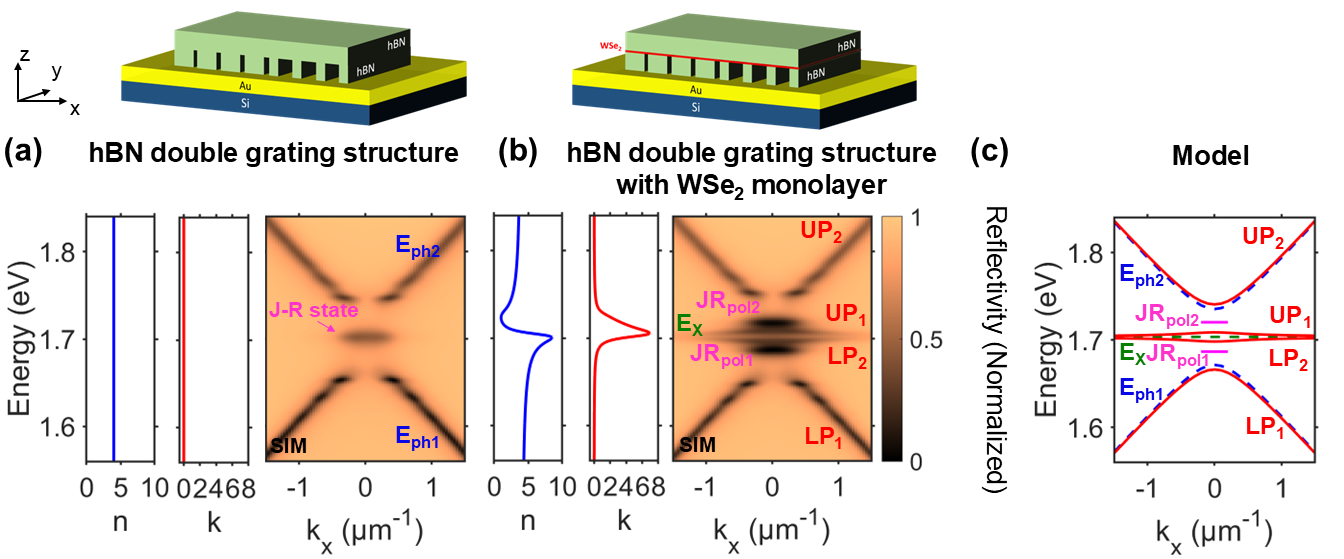}
\caption{\textbf{RCWA \cite{Moharam1981} simulations of double grating hBN heterostructures with and without the influence of a \ch{WSe_2} monolayer exciton.} \textbf{(a)} Bare hBN double-grating heterostructure and corresponding angle-resolved reflectivity, showing a photonic Jackiw-Rebbi (JR) edge state within the band gap. The complex refractive index used in the simulations is shown to the left of the reflectivity map.  \textbf{(b)} hBN double-grating heterostructure containing an encapsulated \ch{WSe_2} and corresponding angle-resolved reflectivity. The excitonic resonance is modelled using a Lorentz oscillator, whose complex refractive index is shown to the left of the reflectivity map. \textbf{(c)} Analytical coupled-oscillator model comparing the bare photonic modes with the hybridised exciton-polariton modes. Strong coupling between the WSe\(_2\) exciton and the photonic modes gives rise to the upper and lower polariton branches and to the two JR-polariton branches, \ch{JR_{pol1}} and \ch{JR_{pol2}}.}
 \label{SI1}
\end{figure*}

Here, we present additional insight into the formation of Jackiw-Rebbi(JR) exciton polaritons in the hBN/\ch{WSe_2} heterostructure. To isolate the purely photonic response, the \ch{WSe_2} monolayer was retained in the RCWA geometry but modelled using a constant background refractive index, without the excitonic resonance. Figure~\ref{SI1}(a) shows the resulting photonic dispersion, including the JR interface state within the bandgap formed by \ch{E_{ph1}} and \ch{E_{ph2}}, as described in Ref \cite{Randerson2026Topological}. 

To incorporate the optical response of the \ch{WSe2} monolayer into the RCWA calculations, its excitonic resonance was modelled using a Lorentz oscillator \cite{Arora2015}. The monolayer was represented as a 0.6 nm thick dielectric layer, with a temperature-dependent dielectric function given by

\begin{equation}
    \varepsilon(E,T) =
    \varepsilon_{\infty} + \varepsilon_X(E,T),
    \label{eq:SI_Lorentz}
\end{equation}

where $\varepsilon_{\infty}=16$ is the background dielectric constant and $\varepsilon_X(E,T)$ represents the excitonic contribution.

The exciton resonance energy was modelled using Varshni's law, while the oscillator amplitude and damping parameter were adjusted to account for their temperature dependence. The resulting complex refractive index, $\tilde{n}(E,T)=\sqrt{\varepsilon(E,T)}$, was incorporated into the RCWA calculations, as illustrated in Fig.~\ref{SI1}(b).

Adding the Lorentz oscillator leads to the hybridisation between the JR edge state, the photonic bands, and the \ch{WSe_2} exciton. The original JR resonance now splits into two branches (\ch{JR_{pol1}} and \ch{JR_{pol2}}). In addition, the upper and lower polariton branches emerge from the dispersive photonic modes. These anticrossing features are characteristic signatures of the strong coupling regime \cite{Xiong2024}.

Figure \ref{SI1} (c) shows the analytical coupled oscillator model used in the main text. The model reproduces the hybridised polariton dispersion obtained from the RCWA calculations, including the splitting of the photonic JR state.

\clearpage
\section{Supplementary Note II}
\begin{figure*}[ht!]
\centering
  \includegraphics[width=\linewidth]{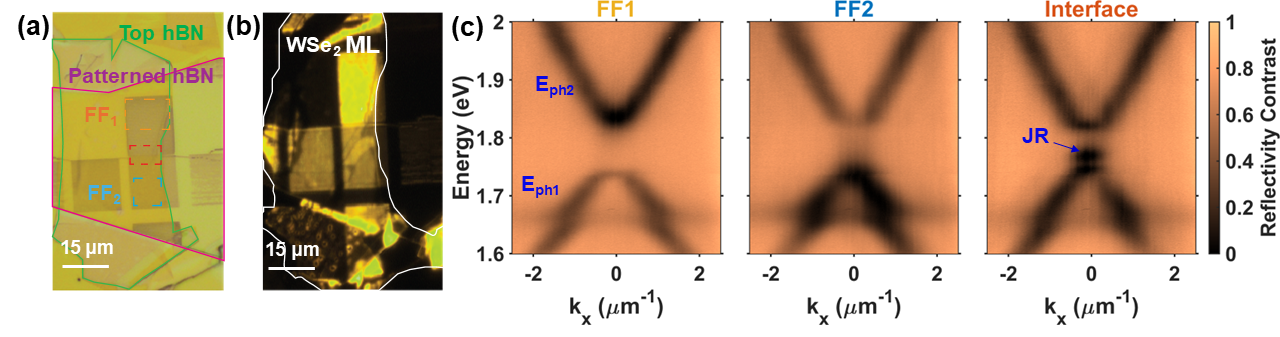}
\caption{\textbf{Room temperature angle-resolved reflectance contrast measurements of an additional hBN/\ch{WSe_2} double-grating heterostructure.} \textbf{(a)} Optical microscope image of the hBN/\ch{WSe_2} double grating heterostructure. The bottom grating layer (86 nm thick) is highlighted in purple, while the top hBN slab (72 nm thick) is highlighted in green. The high and low filling factor regions (FF1 and FF2) and their interface are indicated by the orange, blue and red dashed boxes, respectively.  \textbf{(b)} Photoluminescence image of the same double grating heterostructure. The \ch{WSe_2} monolayer is outlined in white. \textbf{(c)} From left to right: angle-resolved reflectance contrast taken from the high FF, low FF and interface regions, respectively.}
 \label{SI2}
\end{figure*}

To demonstrate the reproducibility of obtaining JR states using the double-grating inverted design, we fabricated and characterised an additional heterostructure. The optical characterisation of this device is shown in Figure \ref{SI2}.

Figures \ref{SI2}. (a) and (b) show optical and photoluminescence images of the heterostructure, respectively. The two grating regions have parameters a1= 470 nm, FF1= 0.73 and  a2= 550 nm, FF2= 0.25.

Figure \ref{SI2} (c) presents the angle-resolved reflectance contrast for the three different regions highlighted. The left panel shows the dispersion for FF1, where a Bound State in the Continuum (BIC)\cite{Hsu2013} resides on the lower photonic band, while a lossy mode occupies the upper band. In the low-FF region, the ordering of these modes is reversed, demonstrating photonic band inversion between the two grating regions. At the interface, an additional resonance appears in the photonic band gap approximately at $1.78\text{ eV}$, corresponding to the JR interface state. Despite the variation in grating dimensions, the inverted grating design again supports a localised JR interface state, confirming the reproducibility of the interface mode in this system.

\clearpage
\section{Supplementary Note III}
\begin{figure}[ht!]
    \centering
    \includegraphics[width=\linewidth]{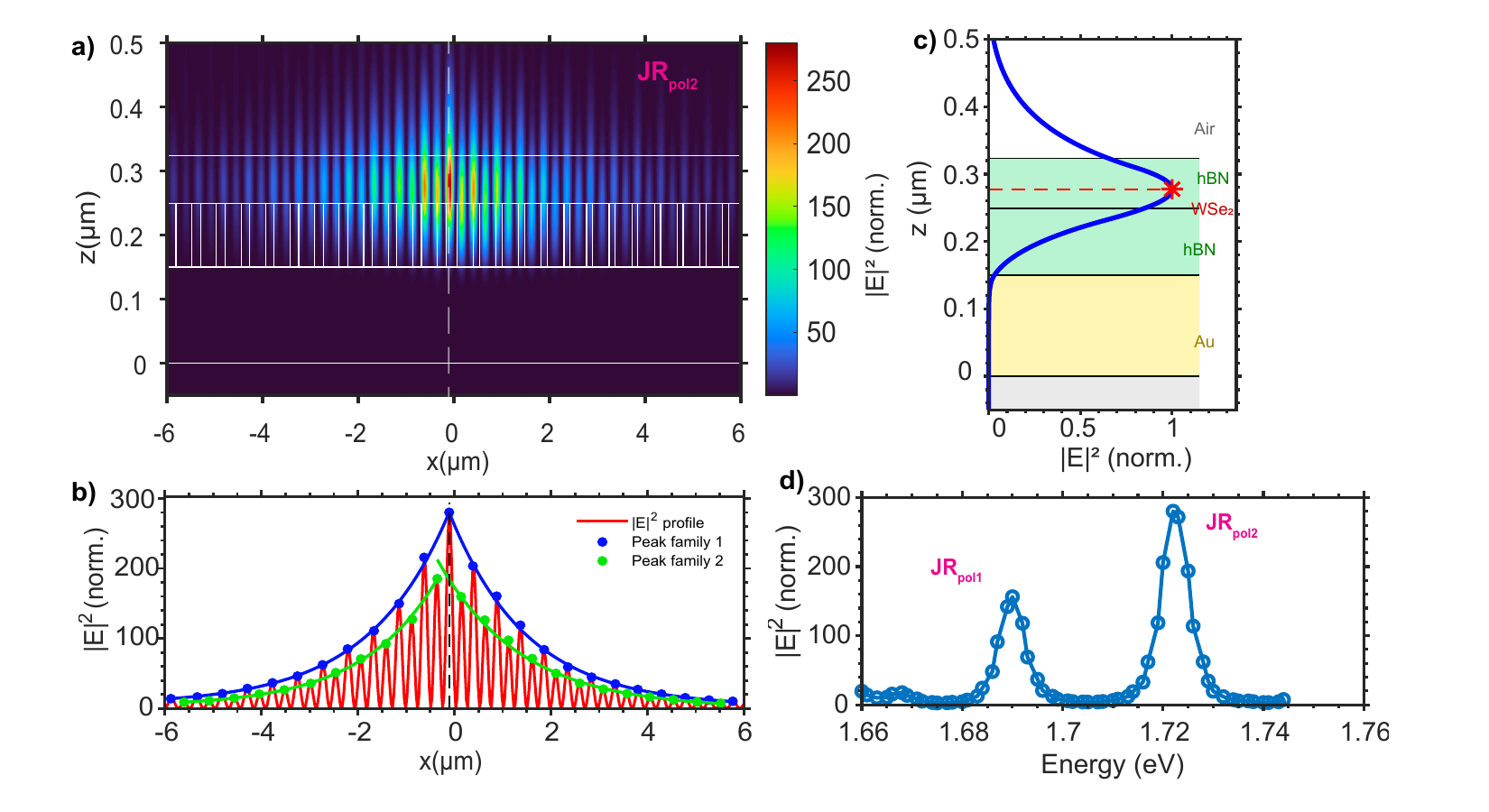}
    \caption{ \textbf{ RCWA-calculated electric field distribution within the inverted hBN/\ch{WSe_2} double grating heterostructure.} \textbf{(a)} Two-dimensional map of the total electric field intensity, calculated at the Jackiw-Rebbi edge state resonance energy. The Grating geometry is overlaid in white. \textbf{(b)} X-plane electric-field intensity showing strong localisation of the mode at the interface. The blue and green lines correspond to the exponential fits performed at either side of the interface. \textbf{(c)} Vertical field profile extracted close to the interface, showing the field distribution across the heterostructure.\textbf{(d)} Maximum electric field at the grating interface for different incident wavelengths.}
    \label{Si2fig}
\end{figure}
Figure \ref{Si2fig} presents a closer look at the RCWA-calculated electric field distribution at the Jackiw-Rebbi polariton resonance. Figures \ref{Si2fig} (a-c) were calculated at an incident wavelength of 720 nm, corresponding to the upper JR-polariton resonance \ch{JR_{pol2}}. The plotted intensity is the total electric-field intensity normalised to the incident wave, $|E|^2/|E_0|^2$.

Figure \ref{Si2fig} (a) and (b) show strong localisation of the mode around the interface between the two band-inverted grating regions. The interface state exhibits a maximum field-intensity enhancement of approximately 294. The lateral field profile decays exponentially away from the interface, with fitted decay lengths of 1.66  $\mu m$ and 1.61 $\mu m$ for the low and high FF areas, respectively. This confirms the confined nature of the JR interface state. The shorter decay length in the high FF area corresponds to the larger effective refractive index in the grating area, resulting in stronger confinement and, subsequently, faster decay in the JR edge state.

The vertical field profile in Figure \ref{Si2fig} (c)  shows that the field maximum lies primarily within the upper hBN slab. This behaviour is consistent with the higher effective refractive index of the slab compared with the grating region. Importantly, a significant field intensity is still present at the position of the encapsulated monolayer, providing the spatial overlap required for strong coupling between the JR edge state and the monolayer exciton.

Figure \ref{Si2fig} (d) shows the maximum interface field intensity as a function of energy. Two pronounced peaks are observed corresponding to \ch{JR_{pol1}} and \ch{JR_{pol2}}. The presence of two confined interface modes, rather than a single photonic JR resonance, confirms the hybridisation of the JR edge state with the WSe$_2$ exciton.

\clearpage
\section{Supplementary Note IV}
\begin{figure*}[ht!]
\centering
  \includegraphics[width=\linewidth]{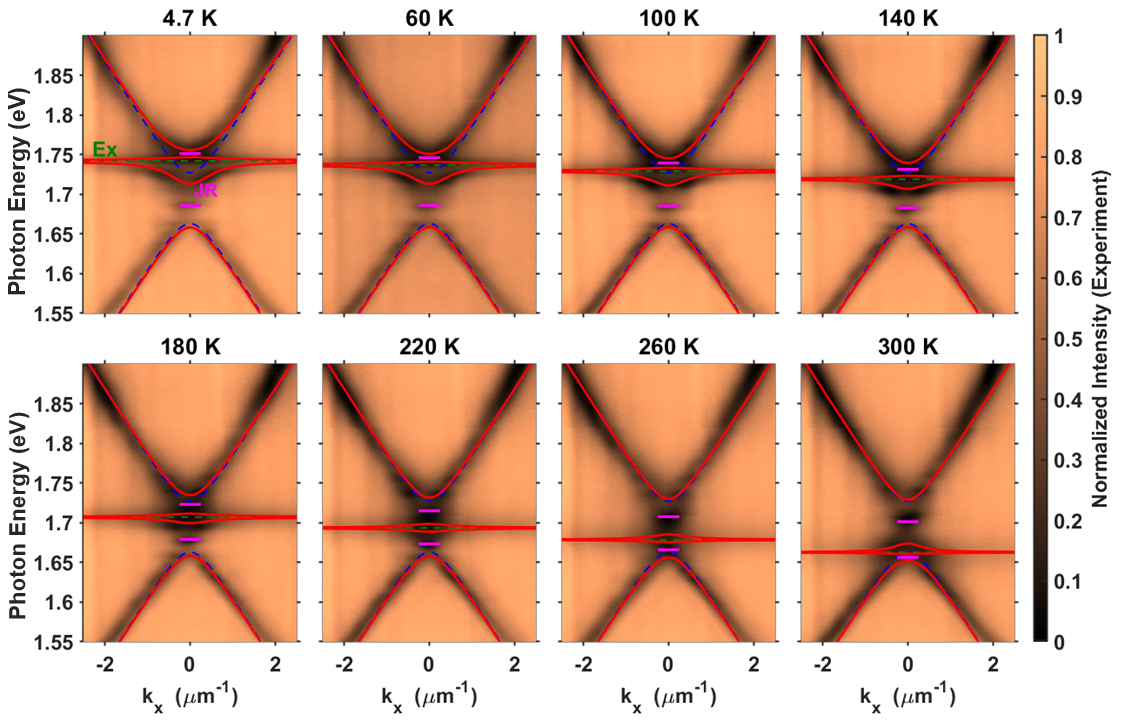}
\caption{\textbf{Temperature-dependent reflectance contrast of the hBN/\ch{WSe_2} topological heterostructure.}  Angle-resolved reflectance contrast spectra of the heterostructure interface measured at temperatures ranging from $4.7\text{ K}$ to $300\text{ K}$. The blue dashed curves indicate the bare photonic modes, the green dashed curve the WSe\(_2\) exciton resonance, the red curves the dispersive polariton branches, and the magenta curves the JR-polariton branches. As the exciton redshifts with increasing temperature, it is tuned successively through the upper photonic band, the JR interface state, and the lower photonic band.} 
 \label{SI3}
\end{figure*}

Figure \ref{SI3} presents further temperature-dependent angle-resolved reflectance spectra of the interface region, complementing the representative temperatures shown in the main text. The exciton, JR edge states, and polaritonic bands are described using the analytical model introduced in Figure \ref{SI1} (c). The bare photonic modes are shown by blue dashed curves, the exciton resonance by the green dashed curve, the dispersive polariton branches by red curves, and the JR-polariton branches by magenta curves. As the temperature increases, the exciton energy is tuned from $1.745\text{ eV}$ at $4.7\text{ K}$ to $1.67\text{ eV}$ at $300\text{ K}$, following Varshni's law \cite{Varshni1967, Arora2015}. This shift in the exciton energy drives the evolution of the strong coupling within the system. At low temperatures, the splitting of the upper polariton band is clearly resolved. As the exciton energy redshifts with increasing temperature, it moves into resonance with the JR edge state, producing a pronounced splitting of the edge state mode. Finally, as the heterostructure is heated further toward room temperature, the exciton tunes into resonance with the lower photonic band, leading to the clear splitting of the lower polariton band.

As shown in Figure \ref{SI3}, the analytical model introduced in Figure \ref{SI1}(c) reproduces our experimental dispersion throughout the temperature range. By adjusting the model parameters to account for temperature-dependent shifts, we accurately captured the system's evolution. For visualisation, the dominant momentum-independent excitonic contribution was numerically removed from the reflectance-contrast maps to improve the visibility of the dispersive photonic and polaritonic modes. The same processing was applied to the corresponding dispersion maps shown in the main text.
\clearpage
\section{Supplementary Note V}
\begin{figure*}[ht!]
\centering
  \includegraphics[width=\linewidth]{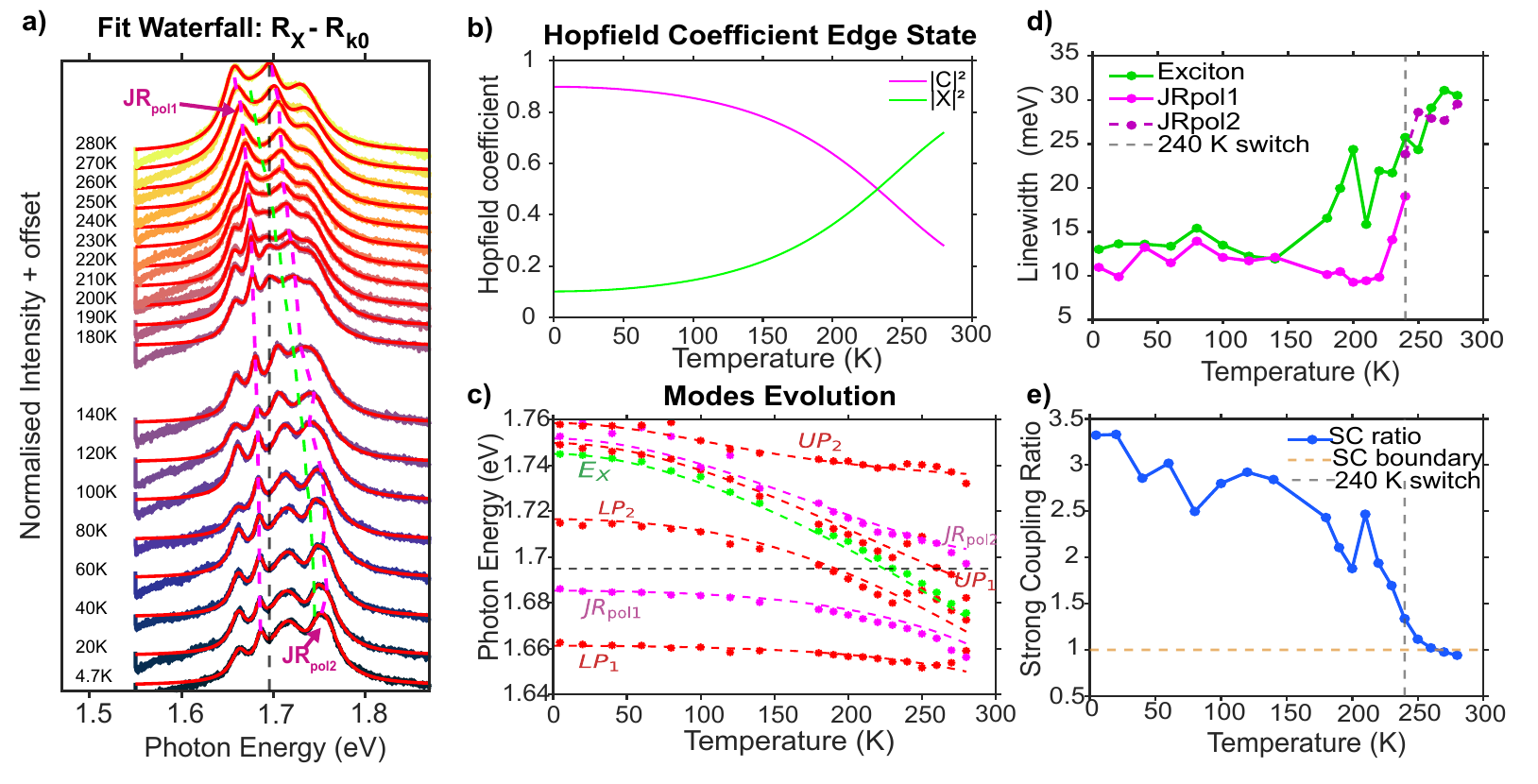}
\caption {\textbf{Experimental evidence of strong coupling between the \ch{WSe_2} exciton and the JR edge state.} \textbf{(a)}$k_x$=0 $\mu m^{-1}$ cross section fitted and experimental trend across all temperatures. The exciton and edge states are highlighted in green and magenta, respectively. \textbf{(b)} Hopfield coefficient of the JR$_{pol1}$ across all temperatures. The photonic factor is highlighted in magenta, while the excitonic in green. \textbf{(c)} Summary of the cross-section fittings showing the trends of the 7 modes present in our system across different temperatures.\textbf{(d)} Linewidth extracted from the fitting for the photonic-like edge states and the excitonic features across all temperatures. \textbf{(e)} Strong coupling ratio as a function of temperature.}
 \label{SI4}
\end{figure*}
To quantitatively assess the coupling between the \ch{WSe_2} exciton and the JR interface state, we analysed the temperature-dependent reflectance spectra measured at the interface. While clear splitting can be observed for the photonic modes at cryogenic temperatures, the focus here is on the splitting of the Jackiw-Rebbi edge states, \ch{JR_{pol1}} and \ch{JR_{pol2}}. The analysis is performed at $k_x = 0$ $\mu\text{m}^{-1}$. By placing the heterostructure over gold, the grating mode resonances change shape from Fano to Lorentzian, allowing us to use a multi-Lorentzian fit to track the evolution of the system. Figure \ref{SI4} shows the resulting waterfall spectra and fits. Six coupled resonances are tracked: \ch{E_{JR_pol1}}, \ch{E_{JR_pol2}}, \ch{UP_{1}}, \ch{UP_{2}}, \ch{LP_{1}}, and \ch{LP_{2}}. The exciton energy, extracted independently from momentum regions where the grating modes are absent, is shown in green. The uncoupled photonic JR state energy is shown by the black dashed line, while the coupled edge states are highlighted with magenta dashed lines.

Figure \ref{SI4} (b) shows the Hopfield coefficients of \ch{JR_{pol1}}. The photonic and excitonic fractions satisfy \(|C|^2+|X|^2=1\). As the temperature increases, JR$_pol1$ evolves from predominantly photonic to predominantly excitonic character.
Figure \ref{SI4}(c) summarises the peak positions, extracted from the fitting, of all modes across all temperatures. The two JR-polariton branches exhibit an avoided crossing around the exciton and bare JR as the exciton resonance is tuned.  Regarding the parameter extraction from the waterfall plot fitting, Figure \ref{SI4}(d) summarises the exciton and photonic (edge state) linewidths across all temperatures. As expected, as the temperature is raised, the exciton linewidth broadens, resulting in a weaker coupling strength. To estimate the photonic damping rate entering the strong-coupling criterion, the linewidth of the predominantly photonic JR-polariton branch was used. Below approximately 240 K, this corresponds to \ch{JR_{pol1}}, while above the Hopfield-coefficient crossover, the predominantly photonic branch is \ch{JR_{pol2}}. The 240 K crossover therefore marks the change in which JR-polariton branch is used to estimate the photonic linewidth, rather than the transition out of the strong-coupling regime.

The Rabi splitting ($\hbar\Omega_R$) was obtained using the model introduced in \ref{SI1}(c). We define the strong coupling ratio as \[R_{\mathrm{SC}}=\frac{\hbar\Omega_R}{(\gamma_{\rm ex}+\gamma_{\rm ph})/2}\], where \(\gamma_{\rm ex}\) and \(\gamma_{\rm ph}\) are the excitonic and photonic linewidths respectively. Strong coupling corresponds to \(R_{\mathrm{SC}}>1\) \cite{Sun2021StrongCoupling}.

As shown in Figure \ref{SI4}(e), \(R_{\mathrm{SC}}\) remains above unity from cryogenic temperature up to approximately 260 K. With increasing temperature, exciton broadening and the reduction of the fitted coupling strength progressively decrease \(R_{\mathrm{SC}}\). Around 270 K the ratio falls below unity, indicating that the system no longer satisfies the strong-coupling criterion used here.

\clearpage
\section{Supplementary Note VI}

\begin{figure*}[ht!]
\centering
  \includegraphics[width=\linewidth]{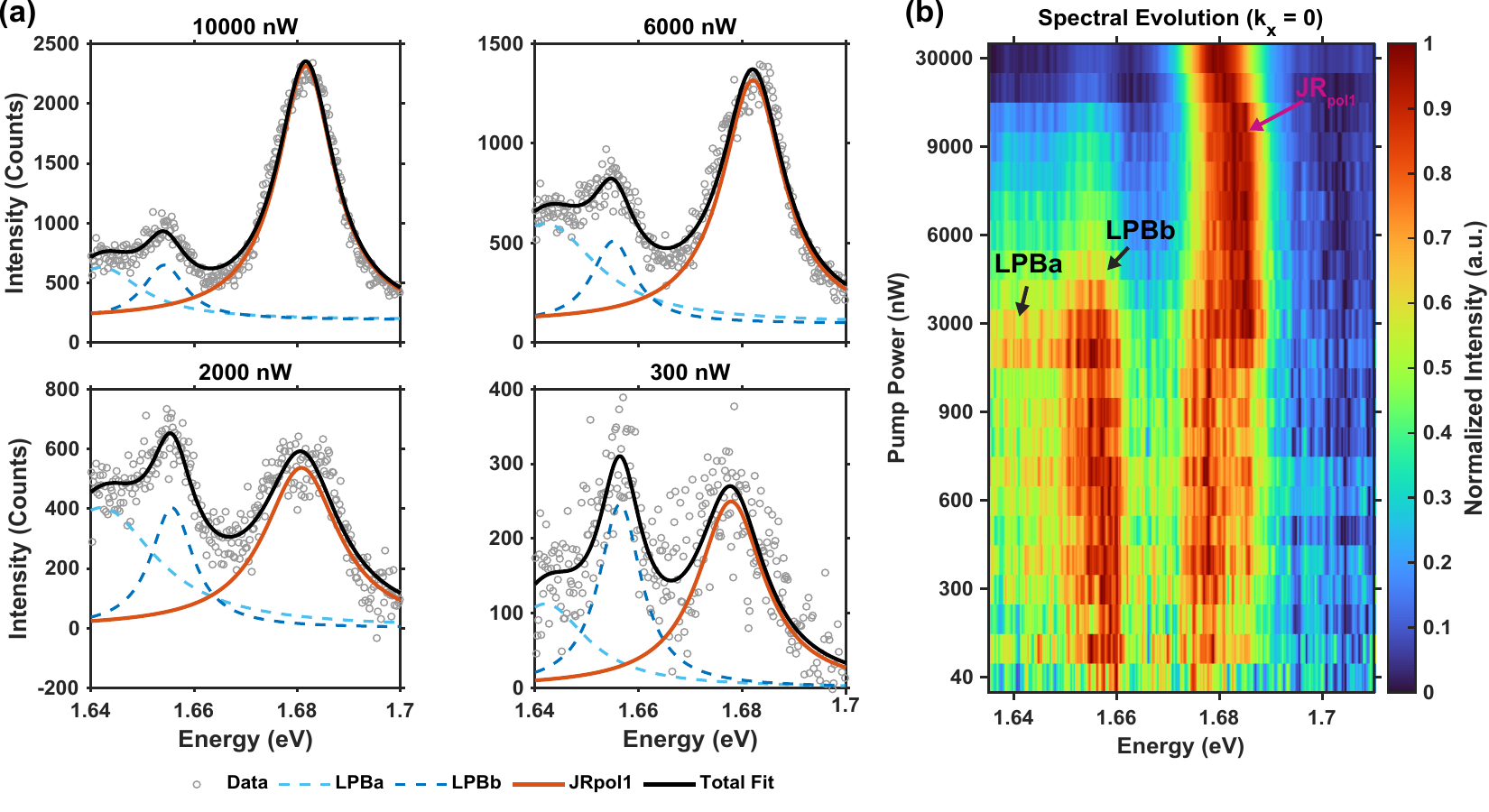}
\caption {\textbf{Photoluminescence power dependence spectral analysis of the Jackiw-Rebbi edge state} \textbf{(a)} Individual PL spectra (circles) and corresponding multi-Voigt fits (solid and dashed lines) at $k_x$ = 0$\mu m^{-1}$ for representative pump powers.\textbf{(b)} Normalised PL intensity heatmap as a function of energy and pump power at $k_x$ = 0$\mu m^{-1}$.} 
 \label{SI6}
\end{figure*}
To quantify the power-dependent evolution of the JR-polariton emission, photoluminescence was extracted around  $k_x$ = 0$\mu m^{-1}$. and fitted for each excitation power. Representative spectra and fits are shown in Figure \ref{SI6} (a). 

The emission spectra were analysed using a multi-Voigt fitting model to distinguish between the various polaritonic components. Each spectrum was decomposed into two lower polariton bands ($LPB_a$ and $LPB_b$) and the JR edge state (\ch{JR_{pol1}}). $LPB_a$ and $LPB_b$ originate from LP1 and arise from the slightly different LP1 resonance energies supported by the two individual grating regions. This fitting procedure allows us to track the evolution of the edge state emission independently of the photonic modes. From the fitted JR component, we extract the peak energy, full width at half maximum (FWHM), and integrated intensity at each excitation power. These quantities are used to construct the power-dependent trends shown in Figure 5(c-e) of the main text. The representative fits in Figure \ref{SI6} (a) illustrate the evolution of the JR resonance across the investigated power range, while the heatmap in Figure \ref{SI6} (b) provides an overview of the corresponding spectral evolution.

\putbib[bib1]
\end{bibunit}

\end{document}